# Finite-thickness effect and spin polarization of the even-denominator fractional quantum Hall states

Pengjie Wang[1,2], Jian Sun[1], Hailong Fu[3], Yijia Wu[1], Hua Chen[1], L. N. Pfeiffer[4], K. W. West[4], X. C. Xie[1,5,6], Xi Lin[1,5,6*]

1 International Center for Quantum Materials, Peking University, Beijing 100871, China
2 Department of Physics, Princeton University, Princeton, New Jersey 08544, USA
3 Department of Physics, The Pennsylvania State University, University Park, Pennsylvania 16802, USA
4 Department of Electrical Engineering, Princeton University, Princeton, New Jersey 08544, USA
5 Beijing Academy of Quantum Information Sciences, Beijing 100193, China
6 CAS Center for Excellence in Topological Quantum Computation, University of Chinese Academy of Sciences, Beijing 100190, China
* xilin@pku.edu.cn

*The spin-polarized even-denominator fractional quantum Hall (FQH) states in the second Landau level (LL), i.e. 5/2 and 7/2, may possess novel quasi-particle excitations obeying non-Abelian statistics. However, the spin polarization of the 7/2 FQH state has not been investigated experimentally and the spin polarization of the 5/2 FQH state from tilted field experiments remains controversial. Using a piezo-driven sample rotator with the lowest electron temperature down to 25 mK, we studied the energy gap of the even-denominator FQH states in the second LL by precise control of the tilted angles with a resolution less than 0.1°. We observed two different energy gap dependences on the in-plane magnetic field for 5/2, 7/2, other FQH states (7/3 and 8/3) in the second LL and reentrant integer quantum Hall (RIQH) states in the third LL. Though the transition fields vary from states, their corresponding in-plane magnetic lengths are comparable to the quantum well thickness of the sample, which may result from the influence of the finite-thickness effect. At low in-plane magnetic fields, before the conjectured finite-thickness effect starts to dominate, the energy gaps of both 5/2 and 7/2 states show a non-decreasing behavior, supporting spin-polarized ground states. Our results also suggest that the 7/3, 8/3 FQH states, and the RIQH states in the third LL are spin-polarized or partially spin-polarized.*

Even-denominator fractional quantum Hall (FQH) states have been focused for years as candidates for demonstrating non-Abelian statistics [1–3]. However, directly probing the statistics is experimentally challenging. For example, although fractional statistics and anyons have been predicted in the two-dimensional electron gas (2DEG) for many years [4,5], there is direct experimental evidence of fractional charge but the proof of statistics itself is rare [6,7]. Some of the predicted wave-functions for the 5/2 and 7/2 even-denominator states [8] are non-Abelian, and they could be supported through other properties, such as the strength of the interaction between quasiparticles [9–11] or the status of spin polarization [12–14]. Spin polarization has facilitated the understanding of complex quantum systems for its unique contribution to choose theoretical models. For example, the paramagnon and Vollhardt models were considered in the spin-polarized liquid $^3$He [15,16]. The property of spin polarization was also well studied in two-dimensional system more than just probing the FQH states' wave-functions, such as the studies of 0.7($2e^2$/h) anomaly [17], metallic state in two-dimensional holes [18,19], non-equilibrium transport [20], zero-field Hall coefficient anomaly [21], confined geometry [22–25] and Coulomb drag [26]. It's proposed that a non-Abelian 5/2 state is spin-polarized theoretically [27]. The spin polarization at 5/2 has been under intense studies through different experimental techniques, such as optical method [28,29], resistive detected nuclear magnetic resonance (RDNMR) [30,31], geometric resonance [32] and tilted field experiments [33–38]. Optical experiments supported an unpolarized 5/2 state, but RDNMR and geometric resonance measurements supported a spin-polarized 5/2 state.

Spin polarization at 5/2 has been investigated by different groups with tilted field experiments but the conclusion remains controversial [33–38]. The status of the spin polarization was determined by how the energy gap of the FQH states at a fixed perpendicular magnetic field changes with the in-plane magnetic field [39,40]. The changing in-plane field and the fixed perpendicular field determine the total magnetic field thus determining the Zeeman energy, keeping the filling factor unchanged. Different from the optical and RDNMR experiments, tilted field experiments need to take the finite-thickness effect into account [41,42]. In experiments, it has been observed that the finite-thickness effect causes the variation of effective mass and Landé $g$-factor in a 2DEG [43]. Calculations [41] examined the wave-function overlap, as a function of quasi-2D layer thickness, between the exact ground state of a model Hamiltonian and the consensus variational ansatz wave functions, and suggested 5/2 and 7/3 states are stable with a finite thickness. The calculation also suggested that 11/5 state is more robust in a pure two-dimensional condition, which contradicts the tilted field experiments [44], but can be understood by the localization of the nearest quantum Hall state [44]. Therefore, the finite-thickness effect varies from different states. In a simplified picture with a given quantum well thickness 2DEG, for some states like 5/2 and 7/3, if the magnetic length induced by the in-plane field is much smaller than the quantum well thickness of a 2DEG, we could expect an ideal two-dimensional condition does not exist anymore. The stability of the 5/2 and 7/2 states should also be influenced by the finite-thickness effect if a high-angle tilted field is applied to a 2DEG. Although the finite-thickness effect is important in the tilted field experiments at 5/2, it has not been carefully examined. In addition, the investigation of either the finite-thickness effect or the spin polarization of the 7/2 state has never been qualitatively reported, so the conclusions on the 5/2 state have never been confirmed by its particle-hole symmetry state.

Experimentally, there are several approaches to generate an in-plane magnetic field. Vector magnets have been commercially available but the in-plane field is limited [45]. Magnet rotation systems require special designs and are inconvenient for cryogenic environments [46,47]. The most widely-used approach is to rotate the sample in a superconducting solenoid magnet. A rotating sample holder can be realized by a worm-gear mechanism [48–50], pressurized liquid $^3$He [51] or piezo-driven method [52–54]. Worm-gear and pressurized liquid $^3$He methods are difficult to precisely and reproducibly control the rotation angle, compared with the piezo-driven method.

Using a piezo-driven rotation system with angle reliability less than 0.1° [53], we measured the energy gap of even-denominator FQH states, conventional FQH states, and reentrant integer quantum Hall (RIQH) states as a function of the in-plane magnetic field. A transition of energy gaps in the order of 1 T, which may originate from titled field influence of the finite quantum well thickness (28 nm), is observed. Before the conjectured finite-thickness effect influences the 2DEG, the energy gaps of the 5/2 and 7/2 states increase with the total magnetic field. Our observation provides evidence for the spin-polarized or partially spin-polarized nature of the 7/2 state.

The experiments were conducted on a high-quality GaAs/AlGaAs sample of van der Pauw geometry, with a quantum well thickness of 28 nm. The measured density is $3.2 \times 10^{11}$ cm$^{-2}$ and the mobility is $2.8 \times 10^7$ cm$^2$ V$^{-1}$ s$^{-1}$. The sample was cooled down in a dilution refrigerator equipped with a 9 T magnet. The sample rotation was realized by a piezo-driven in-situ rotation system [53] which reached a base electron temperature of 25 mK. The sample was illuminated by a red LED at 4.5 K with 15 μA for 1 hour before measurements. Standard lock-in technique as shown in Fig. 1(a) was applied with a 17 Hz excitation of no more than 8 nA along <110> direction. The tilted angle is defined as the deviation from the perpendicular field $B_\perp$, as shown in Fig 1(b). The in-plane magnetic field was applied across the current direction.

The longitudinal resistance $R_{xx}$ and the Hall resistance $R_{xy}$ as a function of the magnetic field without any tilting are plotted in Fig. 1(c). The sample was situated on a piezo-driven rotator and the temperature $T_e$ refers to the electron temperature [53]. The 5/2 and 7/2 even-denominator states, 7/3 and 8/3 odd-denominator states, and features of RIQH states have been well developed, indicating an ultra-high mobility sample was cooled down to low enough temperature. The $R_{xx}$-$B$ traces in the upper spin branch at 0° and 45.3° are compared in Fig. 2(a). The RIQH states are labeled as R3a, R3b, R3c, and R3d, following previous literature [55]. The determination of the tilted angles follows previous methods [52,53]. The perpendicular position ($\theta = 0°$) was determined through the resistor positioner, with an uncertainty of 0.1°. All the angles listed here were cross calibrated by the Hall resistance at low magnetic fields and the resistor positioner on the sample rotator. The stability of the tilted angle was within ±0.05°.

At a high tilted angle such as 45.3°, features of FQH states are invisible, where early experiments unveiled a strongly anisotropic phase [56,57]. If we focus on the $R_{xx}$ minimum at filling factor 7/2, the FQH state is the strongest at around 17.5° and becomes weaker with a higher tilted angle (inset of Fig. 2(a)). By varying the temperature and measuring the $R_{xx}$ minimum of an FQH state, the energy gaps can be derived from the Arrhenius equation $R_{xx} \propto \exp(-\Delta/2k_BT)$, as shown in Fig. 2(b) and (c). The energy gaps of the 5/2 and 7/2 states are plotted as a function of the in-plane magnetic field $B_\parallel$ in Fig. 2(d) for comparison, where energy gaps are normalized by the gap of that particular state at zero in-plane field. The energy gaps of the 5/2 and 7/2 states show a non-monotonic $B_\parallel$ dependence, with a transition magnetic field ~ 1 T. Below the transition field, the energy gaps increase with the in-plane magnetic field. Above the transition field, the energy gaps start to drop with the in-plane magnetic field. For the 7/3 and 8/3 states, the energy gaps show two different trends with a transition field ~ 1.75 T, although they monotonically increase with $B_\parallel$. In a single particle picture, the in-plane magnetic length is 25 nm for 1 T and 20 nm for 1.75 T, comparable to the quantum well width 28 nm of our sample. An in-plane magnetic length of 28 nm corresponds to an in-plane magnetic field of 0.84 T, shown as the vertical dashed line in Fig 2(d). It's sensible to speculate that the contrasting dependences of the energy gaps are caused by the magnetic orbital coupling [41,42] originating from finite quantum well thickness. The quasi-particle's size in the even-denominator states is larger, so the 5/2 and 7/2 states are probably more vulnerable in a non-ideal 2DEG, and the transition field might vary from different states. As a result, the energy gaps of the 5/2 and 7/2 states were suppressed much more severely than the 7/3 and 8/3 at high tilted fields.

If the finite-thickness effect exists in FQH states, then it should also affect other electron-electron interacting states in the same 2DEG. Therefore, we carried out the same tilted experiment for RIQH states to verify the influence of such a finite-thickness effect. The RIQH states in the higher LLs are known for their quantization to the nearest integer plateaus in Hall resistance with zero longitudinal resistance [58–62]. Such a transport signature is proposed to be a collectively pinned electron solid [63–66]. In Fig. 3(a), fully developed RIQH plateaus in $R_{xy}$ with the resolution of ±0.03% are shown. At the third LL, there are four RIQH states (labeled as R4a, R4d, R5a, and R5d) quantized to their nearby integer plateaus. The stability of a RIQH state can be characterized by its insulating behavior as a function of temperature. The RIQH states' energy gaps are defined as $R_{xx} \propto \exp(-\Delta/2k_BT)$ here and plotted in Fig. 3(b) as a function of the in-plane field after normalization. The energy gaps are normalized with the energy gap at 0° ($B_\parallel = 0$ T). Resembling the FQH states, two energy gap dependences with the in-plane magnetic field were observed. Checking the slopes of the energy gaps inside and outside the grey color background of Fig. 3(b), the absolute values of slopes outside the grey area are much larger (>250% difference). The energy gaps of the RIQH states decrease significantly with the in-plane field outside the grey area. The physics of the FQH states and the RIQH states are different but they share similar noticeable dependence of the in-plane field, strongly supporting the appearance of the finite-thickness effect. Although the energy gap of spin-polarized states could be independent of the in-plane magnetic field, much smaller Zeeman energy compared with LL gap could also cause the non-sensitivity of the energy gap as a function of the magnetic field. In the second LL, the energy gaps of the RIQH states are much smaller than the conventional FQH states. In our experimental conditions, it is hard for us to quantitatively explore the energy gaps of RIQH states in the second LL. As a result, we use RIQH states at the third LL to verify the influence of the finite-thickness effect. We note that in the calculation [41], the finite-thickness effect varies from different states. For example, the 7/3 and 5/2 states favor a finite thickness, while some FQH states in the lowest LL, e.g., 1/3 and 1/5, are more robust in a pure two-dimensional condition. Our results from both the FQH states (5/2, 7/2, 7/3, and 8/3) and RIQH states (R4a, R4d, R5a, and R5d) are only consistent in the fact that there are two tendencies of energy gap at low tilted angles and high tilted angles. Neither in theory nor with experimental support, there is a critical field for all FQH and RIQH states regarding to the finite-thickness effect. A non-ideal 2DEG should affect different states quantitatively differently, so our simplified picture that the finite-thickness effect is a possible cause is only qualitatively based on a limited number of states and should be examined by further theoretical and experimental efforts.

The spin-polarization status is probed by the energy gap as a function of the total magnetic field, but not the in-plane field [40]. For a spin-unpolarized state, the energy gap should decrease with the increase of the total magnetic field if the perpendicular field remains constant, because the electrons/quasiparticles' spin would become more polarized when the total magnetic field is increased. We plot the energy gaps of the 5/2, 7/2, 7/3, and 8/3 states as a function of the total magnetic field in Fig. 4. For higher tilted angles, all four FQH states' energy gaps may be suppressed by the finite-thickness effect, which affects the 5/2 and 7/2 states more apparently than 7/3 and 8/3 states. At low tilt, where the conjectured finite-thickness effect hasn't suppressed the FQH states, all energy gaps in Fig. 4 increase with the total magnetic field, so the spin polarization of all four states could not be unpolarized. From a study of energy gap dependence on the quantum well thickness [67], the LL mixing at our density and quantum well thickness is potentially invisible. Moreover, we extracted the absolute g-factor values of 0.34, 0.29, 0.71, and 0.52 from the energy gaps of the 5/2, 7/2, 7/3, and 8/3 states at the low field range. In a single particle picture, the commonly accepted value in GaAs is -0.44. The angles where the finite-thickness effect may start to matter in Fig. 4 are less than 20° for four FQH states. Therefore, a piezo-driven rotator with precise angle control and stability is crucial in such a measurement.

In our experiment, we kept the filling factor constant, and increased the total magnetic field by increasing the in-plane field. Such a study cannot be approached through varying the electron density, where a given FQH state can be studied with the same filling factor but different perpendicular magnetic field. It has been known that the quality of 2DEG and the stability of FQH states are partially determined by the density [67]. For example, the mobility changes with the density [68], and also modifies the energy gap of the 5/2 state.

The spin polarization of the 5/2 state under a tilted magnetic field has been explored by several groups [33–36,38]. The decreasing energy gaps at 5/2 are observed in all the works when high tilted angles are reached, sharing the same results with our experiments. Some previous works speculated a spin-unpolarized state while some proposed the spin-polarized possibility. Table I lists the summary of other tilted field measurements at 5/2. At high tilted angles, the in-plane magnetic lengths $l_{//}$ were typically smaller than the quantum well thickness, and the 5/2 state was suppressed [35,38], same as Fig. 2(a) in this work. At low tilted ranges, the works at the 5/2 state in quantum well thickness of 40 nm [35] and 50 nm [38] respectively, reported a narrow range of non-decreasing behavior when the in-plane length was smaller than the quantum well thickness. Our observation for the even-denominator states is consistent with the previous studies: before the finite-thickness effect starts to dominate, the enhancement of energy gap at 5/2 suggests the ground state should be spin-polarized or partially spin-polarized.

Moreover, the features in the upper spin branch ($3 < \nu < 4$) are broadly similar to those in the lower spin branch ($2 < \nu < 3$). As the particle-hole conjugate state of 5/2, the 7/2 state, which emerges at the upper spin branch in ultra-high quality samples, is believed to possess the same properties as the 5/2. The candidates of the ground state at 5/2, should also be the candidates at 7/2. However, there are rare studies of the 7/2 state due to its lower energy gap and thus more challenging experimental conditions. The 7/2 state has only been investigated in wide-well samples [69] and under hydrostatic pressure [70]. The study in wide-well samples [69] explored the evolution of the two-subband 2DEG at and near filling factor 7/2, reveals distinct metamorphose of the ground states. Moreover, evidence of strengthening 7/2 state was observed in the wide-well samples [69], while quantitative studies have been lacking. The hydrostatic pressure study observed a pressure-induced paired-to-nematic transition at 7/2 state, which had also been observed at 5/2 [71,72], indicating a uniform physics of the even denominator states.

Benefited by the ultra-low electron temperature and precise control of our sample rotation system, we found evidence of spin polarization in the 7/2 state with the tilted field method. At low tilted angles, the increasing energy gap indicates the spin-polarized nature at 7/2, supporting the non-Abelian ground state candidates [12,14,27]. The 7/2 state's spin polarization status is consistent with our observation for 5/2 in this work, indicating 7/2 state sharing the same physics as that at 5/2. Whether the non-Abelian statistics exists in even denominator states still requires confirmation from new experimental efforts in addition to the information of spin polarization. We note that the energy gaps of the 7/2 state in our sample is around 120 mK when no in-plane magnetic field is induced. Slightly tilting the sample helps to enhance the energy gap up to ~ 145 mK, which is the largest energy gap of the 7/2 FQH state that ever reported, comparable to the 120 mK from another high-quality measurement [73].

**Acknowledgements:**
We thank D. E. Feldman, Kun Yang, Xin Wan, Zi-Xiang Hu and Yang Liu for discussions. The work at PKU was supported by Beijing Natural Science Foundation (JQ18002), the NSFC (11674009, 11921005), the National Key Research and Development Program of China (2017YFA0303301) and the Strategic Priority Research Program of Chinese Academy of Sciences (Grant No. XDB28000000). The work at Princeton University was funded by the Gordon and Betty Moore Foundation through the EPiQS initiative Grant GBMF4420, by the National Science Foundation MRSEC Grant DMR-1420541, and by the Keck Foundation.

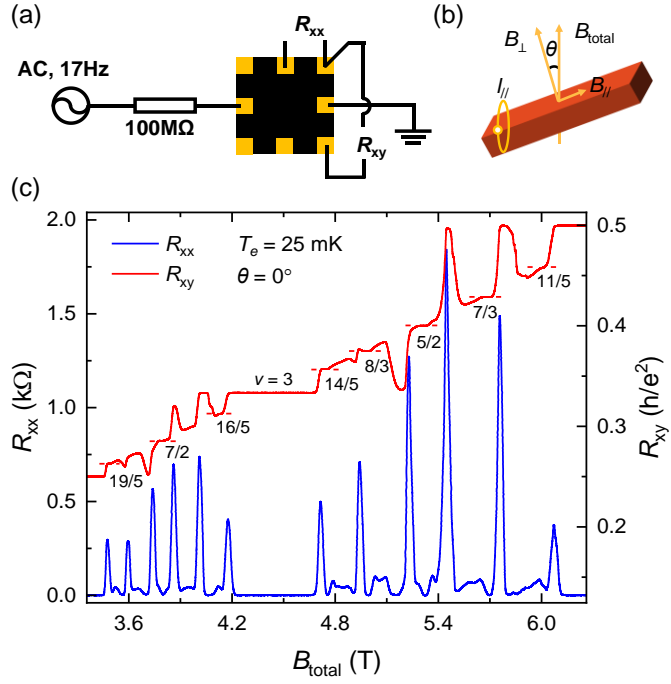

**Figure 1 (a)** Schematic view of the van der Pauw geometry measurement setup. **(b)** Illustration for a sample situated in a tilted magnetic field. The in-plane magnetic length $l_{//}$ is derived from $l_{||} = \sqrt{\frac{\hbar}{eB_{||}}}$, where $B_{//}$ is the in-plane magnetic field. **(c)** Longitudinal (blue line) and Hall (red line) resistances as a function of the perpendicular magnetic field ($\theta = 0°$) in the second LL at 25 mK. The 5/2, 7/2 states and the RIQH states in the second LL only appear in high-quality samples and low electron temperature.

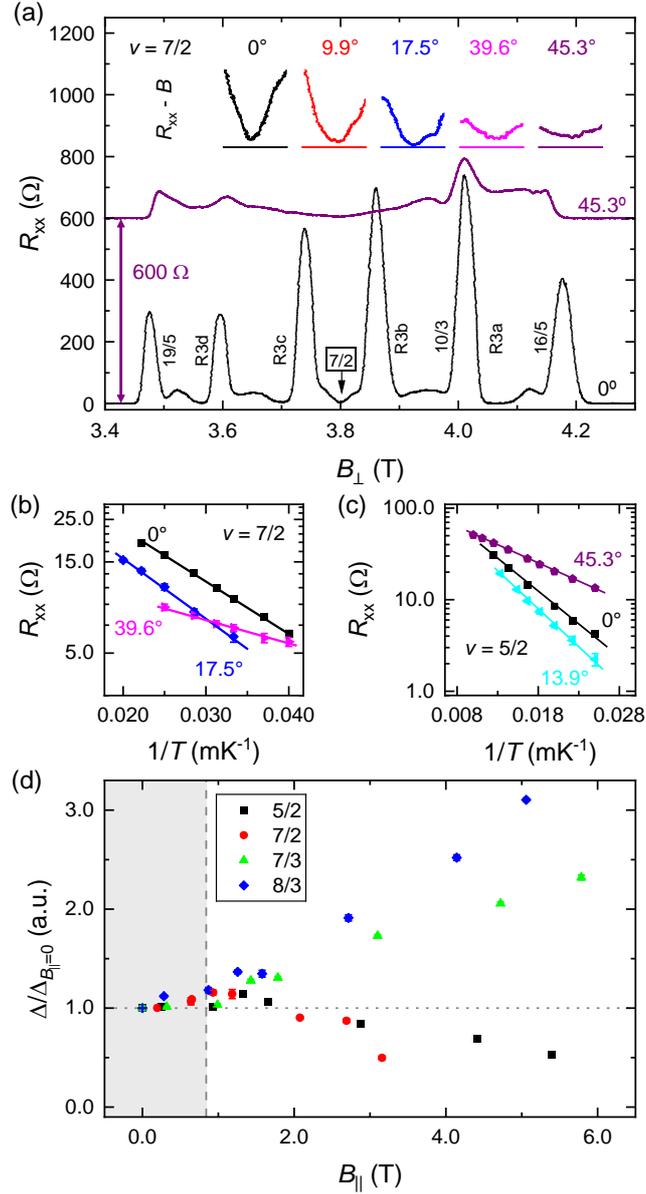

**Figure 2** (a) Longitudinal resistance $R_{xx}$ as a function of the perpendicular magnetic field in the upper spin branch of the second LL ($3 < \nu < 4$) at 25 mK. $R_{xx}$ at 45.3° is shifted by 600 Ω vertically for clarity. The inset shows how the dip of the $R_{xx}$-$B$ curve at 7/2 filling changes with the tilted angle. (b) The temperature dependence of $R_{xx}$ at 7/2. The uncertainty for $R_{xx}$ is typically ±0.3 Ω. The uncertainty for the energy gap ranges from 5% to 10%, depending on the state and tilted angle. (c) The temperature dependence of $R_{xx}$ at 5/2. (d) Normalized energy gaps of FQH state at filling factors 5/2 (black), 7/2 (red), 7/3 (green), and 8/3 (blue) as a function of the in-plane magnetic field. The grey dashed line is for an in-plane magnetic field of 0.84 T.

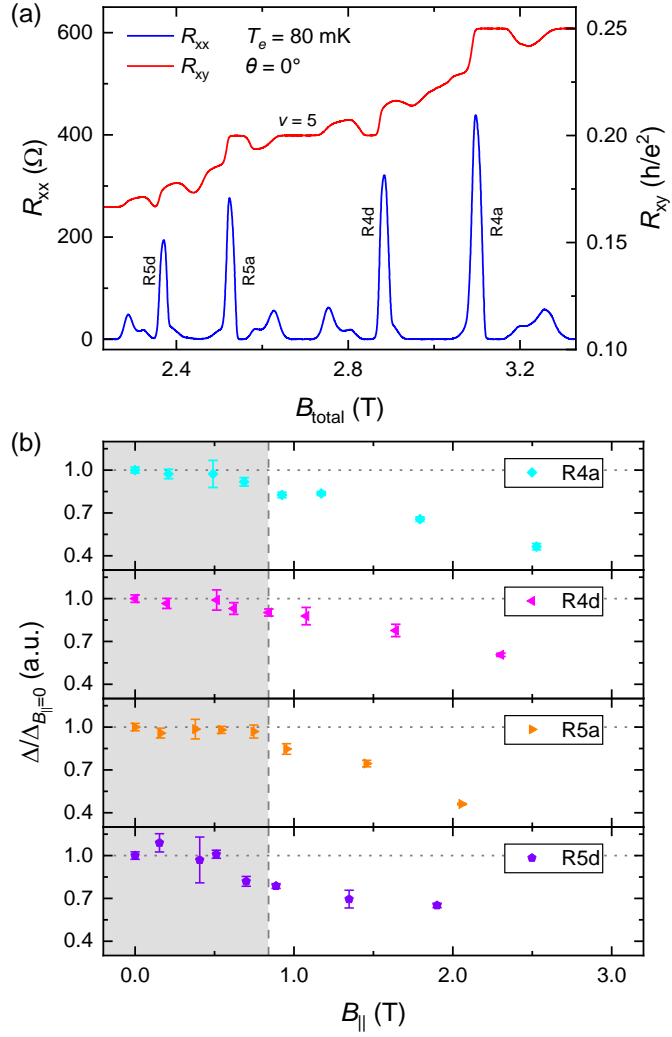

**Figure 3 (a)** Longitudinal (blue line) and Hall (red line) resistances as a function of the perpendicular magnetic field ($\theta = 0°$) in the third LL at 80 mK. **(b)** Normalized energy gaps of RIQH states in the third LL as a function of the in-plane magnetic field. The grey dashed line is for an in-plane magnetic field of 0.84 T.

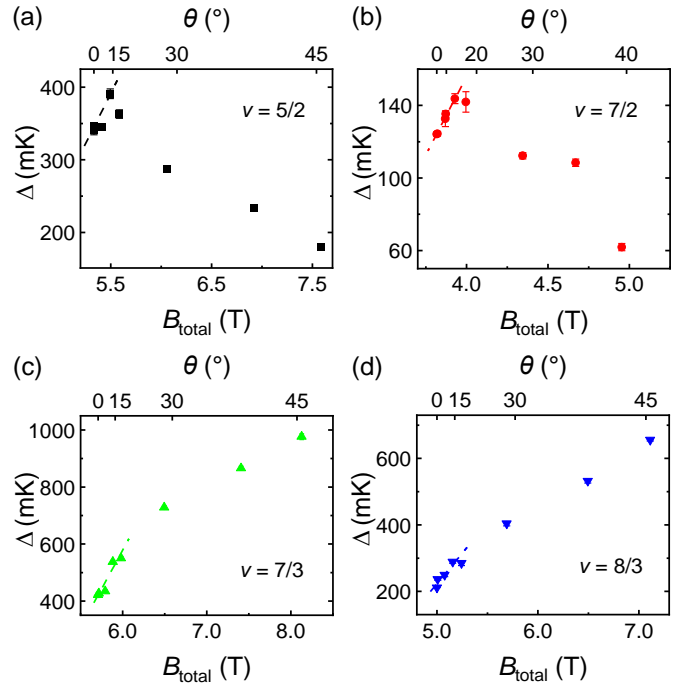

**Figure 4** Energy gaps of the **(a)** 5/2, **(b)** 7/2, **(c)** 7/3 and **(d)** 8/3 FQH states as a function of the total magnetic field. The total magnetic field can be derived from $B_{\text{total}} = B_{//}/\sin\theta$. At low tilted angles, the energy gaps of all four FQH states increase with the total magnetic field. Linear fits (dashed lines) are used to extract the g-factor values from the low-field increasing energy gaps.

| Sample Type | $n_e$ (cm$^{-2}$) | $\mu$ (cm$^2$V$^{-1}$s$^{-1}$) | $\Delta_{5/2}$ (mK) at $\theta = 0°$ | $\Delta_{7/2}$ (mK) at $\theta = 0°$ | $B_{//c}$ (T) | $\theta_{c\,(5/2)}$ | $\theta_{max}$ |
|---|---|---|---|---|---|---|---|
| Heterostructure [33] | $3 \times 10^{11}$ | $0.13 \times 10^7$ | - | - | - | - | 48° |
| Heterostructure [34] | $2.3 \times 10^{11}$ | $0.7 \times 10^7$ | 105 | - | - | - | 18° |
| 40 nm Quantum Well [35] | $1.6 \times 10^{11}$ | $1.4 \times 10^7$ | 262 | 35 | 0.41 | 8.9° | < 40° |
| 40 nm Quantum Well [37] | $1.6 \times 10^{11}$ | $1.6 \times 10^7$ | ~ 215 | - | 0.41 | 8.9° | 76° |
| 20 nm Quantum Well [36] | $6.3 \times 10^{11}$ | $1.0 \times 10^7$ | 125 | - | 1.65 | 9.1° | 49° |
| 30 nm Quantum Well [37] | $1.6 \times 10^{11}$ | $1.6 \times 10^7$ | ~ 145 | - | 0.73 | 16.1° | 77° |
| 50 nm Quantum Well [38] | $1.0 \times 10^{11}$ | $1.5 \times 10^7$ | ~10 | - | 0.26 | 9.0° | 60° |

**TABLE I.** Summary of previous energy gap studies of the 5/2 FQH state under the tilted magnetic field. The transition in-plane field should not exist for heterostructure samples, and references [33] in 1988 and [34] in 1990 supported a spin-unpolarized 5/2 FQH state. The in-plane magnetic field calculated from the quantum well thickness is noted as $B_{//c}$ and the corresponding tilted angle for the 5/2 FQH state is noted as $\theta_{c\,(5/2)}$.